\documentclass[12pt]{iopart}

\usepackage[dvips]{graphicx}

\begin{document}

\title[Reception frequency bandwidth\dots]{Reception frequency bandwidth of a gravitational resonant detector with optical readout}

\author{ A.V.Gusev, V.N.Rudenko, S.A.Cheprasov\dag\\
M.Bassan\ddag}

\address{\dag SAI. MSU, Moscow, Russia;}
\address{\ddag Dip. Fisica, Universit\`a
Tor Vergata and INFN\, Roma, Italy}

\begin{abstract}
A gravitational resonant bar detector with a large scale
Fabry-Perot cavity as an optical read out and a mechanical displacement transformer
is considered. We calculate, in a fully analytical way, the final receiver bandwidth
in which the potential sensitivity, limited only by the bar thermal noise, is maintained
despite the additional thermal noise of the transformer and the additive noise of the optical
readout. We discuss also an application to the OGRAN project,
where the bar is instrumented with a 2m long FP cavity.
\end{abstract}

\submitto{\CQG}
\pacs{0480N,0760}

\maketitle

\section{Introduction}

A search for a new type of cosmic radiation -- gravitational
waves, -- has been carried out during last years with two kinds of
gravitational wave detectors: cryogenic resonant bars \cite{IGEC}
and laser interferometers on suspended mirrors \cite{hpage}. Bars
have already been in operation for more than ten years. Meanwhile
interferometers are beginning their active "scientific life".
 Starting with year 2008 a series of incremental upgrades of the
interferometers is planned. In the next five years there will be long
periods when most of these observatories will be out of operation to
complete such upgrades, while resonant bars will remain in  scientific duty,
accumulating observational data. The great reliability and the low cost
of operation, demonstrated in many years of observations, make it reasonable
and worthwhile to keep resonant detectors on the air as "supernova watch"
during these coming years. Their sensitivity has proven to be accurately
predictable, on the basis of  detector dynamics and fundamental noise sources,
with the possible exception of some excess in the high energy tail of the
event distribution.

A clear advantage of the "free mass interferometer GW detector"\,
consists in its wide detection  frequency band $\sim10^{3}$ Hz.
However this band is not a homogeneous low noise region, being
composed of a number of "windows of good
sensitivity"\, separated by peaks of increased noise
associated with resonances of suspensions and other technical
causes. The typical width of such window is on order of $\sim100\div
300$ Hz. The best sensitivity achieved at present inside the
window in the region a few hundred Hz is characterized by the
noise spectral density $\sim3\cdot10^{-23}\,\mathrm{Hz}^{-1/2}$;
in the zone of the bar vision $(1\pm 0.2) kHz$ noises are
somewhat higher $\sim1\cdot10^{-22} \mathrm{Hz}^{-1/2}$) and
growing linearly with frequency in terms of metric perturbations \cite{hpage}.

The present generation of "cryogenic resonant bar detectors"\, was
designed for  cooling below 100 mK: however they  actually operate
at the temperature $\sim(2\div3)\mathrm{K}$ in the "kHz"\, region
reaching almost the same noise spectral density
$\sim5\cdot10^{-22}$, $\mathrm{Hz}^{-1/2}$ but in a very narrow
frequency band $\sim(1\div0.1)$ Hz.
In recent years, all
cryogenic bar groups have modified their read-out systems in order
to achieve a larger bandwidth ($20\div100$ Hz). This was achieved
with a small loss in peak sensitivity, now in the $
10^{-21} \mathrm{Hz}^{-1/2}$ range (that is  however by no mean
fundamental: it is mainly due to the lower Q factor of high coupling
devices  and could be regained in possible future upgrades).

The limitation of the bar reception bandwidth arises from noises
of the read out, but not from the thermal fluctuation of the bar
itself. Indeed,  the potential sensitivity of the bar cooled down
to the temperature $10\,\mathrm{mK}$ and equipped by a noiseless
read out sensor corresponds to its thermal noise spectral density
$\sim10^{-23}\mathrm{Hz}^{-1/2}$, i.e to the noise level projected
for advanced interferometers at  $1$ kHz. A clear understanding of
this fact stimulates the "bar groups"\, for an investigation of
new types of low noise recording devices. In that number the
fist place belongs to the "optical read out"\, which in fact was
successfully used in the interferometers to measure the
displacements of test mass-mirrors
\cite{hpage}. \par
In this paper we analyze reception characteristics (the sensitivity
and bandwidth) of a gravitational bar detector with mechanical
transducer (displacement transformer) and an optical FP-cavity as a
read out system. As an intermediate step we consider a possibility
of using a displacement transformer as a matching link between the
bar and FP-cavity at room temperature (the Russian project "OGRAN"\,
\cite{cambridge}). However our final target is a calculation of "the
maximum receiver bandwidth"\, for the "super cryogenic bar"\,
($10\, \mathrm{mK}$) with optical read out at the level of its
potential sensitivity. We have carried out our analysis considering as
background noise only that  of unavoidable natural fluctuation: thermal
noises of mechanical parts and shot noise in optical channels. Non fundamental
limitation of technical noises is discussed briefly in a comparison with
experiments performed in recent years.

\section{Equivalent scheme and notations}

All existing gravitational bar detectors utilize the displacement
transformer (DT) - a light mechanical oscillator attached to one
end of the aluminum bar, for an impedance matching of the bar
acoustical resonator to the EM-sensor (which is usually part of a
LC resonance circuit + DC SQUID for cryogenic detectors)
\cite{explorer}, \cite{nautilus}, \cite{cerdonio}. Mechanical
construction of DT was realized in the form of "membrane", "loaded
diaphragm", "mushroom"\, or "rosette". At the resonance condition
$\omega_{\mathrm{Bar}} = \omega_{\mathrm{DT}}= \omega_{0}$ the
energy of bar's oscillation is transferred to DT with the "beating
frequency"\, $\Omega_{B}=0.5\sqrt{\mu} \,\omega_{0}$, where the
small parameter $\mu = {m/M}\ll 1 $ is defined by the ratio of DT
and bar masses. Under a good matching, the  DT amplitude is larger by
the factor $\mu^{-1/2}$ than the bar amplitude.\par
A general theory of the bar with DT  was initially published in
\cite{paik}, \cite{piter} and then elaborated in many papers of "bar
groups". A principal role of the optical read out for the classical
Weber bar was theoretically investigated in the paper
\cite{kulpolrud}. Afterwards, this idea was developed in
\cite{richar} and then a pilot model was designed and tested at room
temperature \cite{richar2}, \cite{richar3}, \cite{mio}. The
possibility to achieve the sensitivity $10^{-20}Hz^{-1/2}$ for a
room temperature bar with a good optical readout have been shown in
\cite{guskulrud}. The correctness of this estimate was demonstrated
in the paper \cite{auriga} where  the first full implementation of a
bar with opto-mechanical
readout  was constructed and tested.  \par
There are at least two ways to couple a bar with an optical
FP-cavity. In both cases one mirror of the cavity must be clamped to
the membrane of DT. For the second mirror one has a choice: it may
be attached either to the front (near) or to the opposite (far)
bar's end surface. For conventional bars \cite{explorer},
\cite{nautilus}, \cite{cerdonio} mainly the front end can be used.
In this variant DT together with FP-cavity presents a single
measuring element - a sensitive accelerometer. To improve the
coupling and to suppress optical noises the FP-cavity gap must be as
small as possible \cite{mio}, \cite{auriga}. A direct action of GW
on such accelerometer produces much smaller perturbation in
comparison with the bar's reaction and so it may be neglected.

A different situation arises if the second mirror can be attached
to the far end of the bar. Such opportunity is indeed available in
the Russian project OGRAN, in which the cylindrical bar detector
has an optical tunnel along its central axis \cite{cambridge}. In
this configuration the "electromagnetic degree of freedom"\, (long
optical FP resonator) has the same geometrical scale as the
acoustical one. It means that a "GW-reaction"\, of the
electromagnetic part of the detector has to be taken into account
on equal footing with the acoustical one. Actually, as it was
shown in the paper \cite{kulpolrud}, a reaction of such
"opto-acoustical"\, gravitational detector contains, in general,
two separate parts: the conventional acoustical response and a new
one, the optical response. However, in a "long wave limit" (when
the GW wavelength  is much larger the bar length) the optical
response of a high finesse FP-cavity is small enough and, in the
first approximation, may be neglected. \par
In our analysis below we will consider namely the second nontraditional
configuration: a bar with a long FP-cavity; one mirror is attached
to the far bar end face, and the other to a DT fastened to the
opposite bar end. Using a reference system with the origin in the
bar center one can present an equivalent scheme
of the opto-acoustical detector as it is shown in $fig.1$. \par
Additional notations used in this figure are the following. \\
$k_{1}=M\omega_{1}^{2}$, $k_{2}=m\omega_{2}^{2}$ --- coefficients
of rigidity, masses and partial frequencies of the bar and DT with
corresponding coordinates $x_{1},\,x_{2}$. Coefficients of losses
are $H_{1}$, $H_{2}$ (we assume a frictional force,  proportional
to velocity of the mass). The eigen frequencies of the "coupled
oscillatory system"\, are defined as
$\omega_{e1}=\omega_{0}-\Omega_{B}$, $\Omega_{e2}=\omega_{0}
+\Omega_{B}$.\par
One can see the first mirror of FP cavity is fixed at the point
$x=0$, the second moving mirror is clamped to the DT mass. The
auxiliary optical components --- laser, beam splitter and photo
receiver, are symbolically shown in \textit{fig.1}. In the
hypothesis of constant laser power $P$ and frequency $\omega_{e}$
such configuration presents the typical scheme of a parametric
read out: the signal of interest being detected through variations
of some parameter, namely --- the length of FP cavity. \par
Arrows represent all forces of our interest: the equivalent GW
action is represented by a "signal force" \, $F_{s}$ ; the
thermal excitation are symbolically shown through "Langevin stochastic
forces" \,$F_{fl1}$, $F_{fl2}$.\par
Having in mind the model of \textit{fig.1} we can easily write down
the equations of motion, which in general describe a dynamics of
three coupled degree of freedom: bar, DT and FP cavity. Below we
will consider a special case of our interest with a resonance
optical pump. In this case, the bar acoustical oscillation produces
only a phase modulation of the output light; its residual amplitude
variation may be neglected \cite{kulrud}. This means that the
dynamical influence of the FP cavity on the bar is absent: there is
no "optical spring"\, effect \cite{virgilio}. Besides, one can omit
also the "back action fluctuation"\, effect if the pump power is
smaller the critical one, which is extremely large $P_{cr}\ll1$ kW
(see Appendix).\par
In this approximation the third equation, describing the FP cavity,
is reduced to a simple factor of conversion of a "DT mechanical
displacement"\, into the "optical output signal". However a
specific electromagnetic or "read out noise"\, has
to be added to acoustical thermal noises.\par
After this remarks we will write down the equations of motion and will
analyze a solution to estimate the detection sensitivity
and receiver bandwidth.

\section{Equations of motion and noise spectrum}

A signal track block diagram is shown in $fig.2$. GW-signal and
thermal fluctuation are represented by their equivalent forces
reduced to the input of the acoustical part of the GW detector
(bar + DT). The output of this part,- a DT coordinate perturbation
$x_{DT}=x_{2}(t)$, produces variations of the FP cavity length
$L(t)= L_{0}+x_{2}(t)+x_{1}(t)$; being $ x_{2}\gg x_{1}$, we
neglect the last term. This length change leads to phase
modulation of the output light, recorded by a heterodyne
photo detector (HPD). The proper noise of such photo detection (just
a "read out noise") has to be modeled in a general SNR balance.\par
In this paper we would like to avoid further detailing of the block
diagram $fig.2$ keeping in the mind different conceivable schemes of
optical demodulation such as the self photo detection type of the
Pound-Drever technique \cite{auriga} or the use of some external
reference optical flux at the heterodyne photo mixer \cite{mandel}.\par
Introducing the differential operator $p=d/dt$ one can write the
equations of motion of the acoustical part in algebraic form.
The losses are taken into account through the well known formalism
of "complex rigidity": $\tilde {k_{i}}=k_{i}+pH_{i}$
\cite{Landau1}.
$$
\left\{
\begin{array}{c}
[Mp^{2}+p(H_{1}+H_{2})+(k_{1}+k_{2})]x_{1}-(k_{2}+pH_{2})x_{2}=F_{s}+F_{fl1},
\\
(mp^{2}+pH_{2}+k_{2})x_{2}-(k_{2}+pH_{2})x_{1}=F_{fl2}
\end{array}
\right.\eqno(1)
$$
A solution of  these equations for the DT coordinate is written as
$$
\left<x_{2}(t)\right>=K_{m}(p)F_{s}(t).\eqno(2)
$$
where
$$
\begin{array}{c}
{\displaystyle
K_{m}(p)=\frac{k_{2}+pH_{2}}{\det(p)},}\\
\det(p)=[Mp^{2}+p(H_{1}+H_{2})+(k_{1}+k_{2})](mp^{2}+pH_{2}+k_{2})-
(k_{2}+pH_{2})^{2}.
\end{array}
$$

For $H_{1}=H_{2}=0$ one has
$$
\begin{array}{c}
\det(p)\propto(p^{2}+\omega_{e1}^{2})(p^{2}+\omega_{e2}), \\
(\omega_{e1}+\omega_{e2})/2=\omega_{0}\approx\omega_{e1},\,\omega_{e2}.
\end{array}\eqno(3)
$$

Having the signal variable $x_{2}(t)$, as a next step one can calculate
the reaction of the resonating FP cavity to variation of
its length $L=L_{0}+x_{2}$. It will be in the form of a phase
modulated transmitted light. Such signal in particular might be
extracted through a heterodyne photodetection procedure
accompanied by the "Mandel photo current"\, noises \cite{mandel}
(We use an approach in which the pump $E_{p}(t)$ is considered as a
classical light field. Its quantum fluctuation is in fact hidden
in the photo current shot noise.) \par
A direct way of carrying out the "signal-noise"\, analysis would
consist in calculation of output spectra densities and estimation of
the SNR in some bandwidth after a matched filter. However, we will
use in what follows the equivalent but more compact method in which
all internal noises of the device \textit{fig.1} are reduced to the
input of the signal track \textit{fig.2}. It is well known that an
estimate of the input SNR corresponds to a SNR after the optimal
filtering procedure. \par
We indicate with  $N_{e}$ for the spectral density of the "read
out noise"\, reduced to the fluctuation of DT coordinate $x_{2n}$.
Then the total input stochastic force can be written as
$$
F_{n}(t)= F_{T}(t) + K^{-1}_{m}(p)\,x_{2n}(t).\eqno(4)
$$
where $F_{T}(t)$ is the equivalent stochastic force responsible
for thermal noises. In a more general approach its spectrum
density can be found through the fluctuation-dissipation theorem
(FDT) \cite{Landau2}. According to this theorem
$\left<|F_{T}(j\omega)|^{2}\right>=2k_B
T\,\mathrm{Im}\left[K_{m}^{-1} (j\omega)/\omega\right]$, where $k_B$ is
the Boltzmann's constant.. So the
total input spectral density is described by the formula
$$
<|F_{n}(j\omega)|^{2}>=2 k_B T\,\mathrm{Im}\,\left[
K_{m}^{-1}(j\omega)/\omega\right] + |K_{m}(j\omega)|^{-2}N_{e}.
\eqno(5)
$$

A "signal influence"\, is given by the well known equivalent
"GW-force"\, perturbing the bar detector in the weak field
approximation
$$
F_{gw}(t)=(1/2)m_{eq}L_{eq}p^{2}h(t)\sim (1/\pi^2)ML\omega_{g}^{2}h_{0}
\eqno(6)
$$
Here $m_{eq}= M/2$ and $L_{eq} = 4 L/\pi^2$; the signal metric
perturbation is supposed in the form of quasi resonance short
pulse: $h(t)=h_{0}\sin{\omega_{g}t},\,\,0<t<\tau_{s}$ so that
$\omega_{g}\sim\omega_{0}$ and $\omega_{g}\tau_{s}\sim 2\pi$, i.e.
a GW-pulse containing only a few periods of carrier, with a duration
much shorter than the relaxation time of the bar. \par
In order to start with SNR analysis we need to define $N_{e}$.

\section{Equivalent optical read out noise}

Let us consider the physics of a FP sensor more in detail in order
to clarify the read-out spectral density $N_{e}$. The three principal
parameters of the cavity are the amplitude coefficients for mirror
reflection $r$, transmission $\tau$ and absorption $a$ which satisfy
the following relations
$$
r^{2}+|\tau|^{2}+a^{2}=1 \, , \tau=j|\tau|.\eqno(7)
$$
The two  FP cavity mirrors are supposed to be equal. \par
Let the complex amplitude of the laser pump be $\tilde{E_{P}}$.
Then in a one mode approximation the amplitude of transmitted
light $E_{\tau}$ is derived by the equation
$$
\begin{array}{c}
(p^{2}+2\gamma p + \omega_{n}^{2})E_{\tau} = 2\gamma_{\tau}pE ;\\
E=2(-1)^{n}E_{P},\;\omega_{n}=\pi(c/L_{0})n,\;\gamma=(\gamma_{\tau}+
\gamma_{a}),\,L=L_{0}+x_{2}
\end{array}
\eqno(8)
$$
Here $\omega_{n}$ is the resonance frequency of $n-th$ optical mode
of the cavity; the relaxation index $\gamma$ (half width of the resonance)
has two contributions: one $\gamma_{\tau}=(c/4L)|\tau|^{2}$ due to light
leaking from cavity; second $\gamma_{a}=(c/4L)a^{2}$ due to light absorption
in the mirrors.\par
In a narrow region near the optical resonance $\omega_{P}\sim\omega_{n}$
the complex amplitude of the transmitted light $E_{\tau}$ can be approximated as
$$
\begin{array}{c}
{\displaystyle
\tilde{E}_{\tau}\sim\frac{j\gamma_{\tau}}{(\Omega+\Delta-j\gamma)}
\tilde{E}(j\omega),} \\
\Omega=\omega-\omega_{n};\,\Delta=\omega_{P}-\omega_{n}\sim
\omega_{n}(x_{2}/L).
\end{array}
\eqno(9)
$$
The tuning at resonance $\Delta=0$, with the initial
phase choice $\tilde{E}_{P}=E_{0}$, provides the signal
amplitude of the transmitted light
$$
\tilde{E}_{\tau}\approx(-1)^{(n+1)}\frac{2\gamma_{\tau}}{\gamma}
E_{0}\exp\left\{-j\frac{\omega_{P}}{\gamma}\,\frac{x_{2}}{L}
\right\}\eqno(10)
$$
Thus the DT coordinate variation $x_{2}(t)$ produces a phase
modulation of the transmitted light. The corresponding optical
perturbation can be detected for example through the heterodyne
photodetection with a reference field $E_{g}$ at the resonance frequency.\par
The only natural source of fluctuation in the approach of classical
pump is a photo current shot noise $i_{n}(t)$
$$
\begin{array}{c}
I(t)=I_{0}+i_{n}(t),\;I_{0}=\eta(eP_{g}/\hbar\omega_{n}),\\
<i_{n}(t)i_{n}(t+\tau)>= eI_{0}\delta(\tau),
\end{array}\eqno(11)
$$
here we assume that the average current $I_{0}$ is
produced mainly by the heterodyne power $P_{g}$; the symbols $e$, $\eta$
are used for the electron charge and the photo detector quantum efficiency.\par
Expanding the exponent (10) for small variation $x_{2}(t)$ one can
get the equivalent current variation
$$
i(t)\approx I_{0}
\frac{2\gamma_{\tau}}{\gamma}\sqrt{\frac{P_{0}}{P_{g}}}\,
\frac{\omega_{n}\,x_{2}(t)}{\gamma\,L}\eqno(12)
$$
where $P_{0}$ is the transmitted light power.  This formula allows us to
consider the fluctuation  $x_{2n}(t)$ as an equivalent "read out
noise"\, putting $i_{t}=i_{n}(t)$. The correlation function of
this noise is
$$
\left<x_{2n}(t)x_{2n}(t+\tau)\right>=\left(\frac{\gamma}{2\gamma_{\tau}}
\right)^2\frac{\hbar\omega_{n}}{\eta P_{0}}\left(\frac{\gamma L}
{\omega_{n}}\right)^{2}\delta(\tau).
$$
After substitution for the parameters $(\gamma,\gamma_{\tau})$ one comes
to the spectral density of the equivalent read out noise
$$
\left<|x_{2n}(\omega)|^{2}\right>=\left(\frac{\lambda}{4\pi}\right)^{2}
\left(|\tau|+\frac{a^{2}}{|\tau|}\right)^{4}
\frac{\hbar\omega_{n}}{\eta P_{0}}.\eqno(13)
$$
A minimum of this noise is achieved under the well known condition from
the theory of FP sensors: the optimal tuning corresponds to
equality of transmitted and absorbed parts of the light
$|\tau|=a$. Finally the spectral density of "read out noise"\,
looks like
$$
N_{e}=\left<|x_{2n}(\omega)|^{2}\right>=
\left(\frac{\lambda}{2\pi}\right)^{2}|\tau|^{4}\,
\frac{\hbar\omega_{n}}{\eta P_{0}}. \eqno(14)
$$
Just this expression has to be used in the formula (5).

\section{SNR analysis}

A knowledge of the signal properties (6) and noise spectrum (5) in
term of equivalent input forces perturbing the opto-acoustical GW
detector allows us to perform a signal-noise analysis in different
conceivable cases.

\subsection{Potential sensitivity}

We shall  call  "potential sensitivity"\, a combination of
amplitudes $F_{\min},\,h_{\min}$ which can be registered in the
"signal bandwidth"\, $\delta\omega=2\pi/\tau_{s}$ by the bar
detector with ideal (noiseless) read out sensor, i.e. only limited
by the bar thermal noise in the entire signal bandwidth. The corresponding
detection condition is
$$
\mathrm{SNR}=\rho_{0}=\frac{1}{4\pi}\int\limits_{-\infty}^{\infty}\frac{|F_{s}
(j\omega)|^{2}}{2 k_B TH_{1}}d\omega = 1. \eqno(15)
$$
For short signal pulses one can use the approximation
$F_{s}(j\omega)\approx F_{0}\tau_{s}$. The substitution in the
integrand (15) leads to the well known formulas
$$
F_{\min}=\sqrt{\frac{2k_B TH_{1}}{\tau_{s}}},\quad
h_{\min}=\frac{2}{L}\left(\frac{k_B T}
{m_{eq}\omega_{0}^{2}}\,\frac{1}{Q\omega_{0}\tau_{s}}\right)^{1/2}
\eqno(16)
$$

\subsection{Sensitivity of the opto-acoustical bar without DT}

We now consider, as an intermediate step toward the full calculation,
the set-up in which the displacement transformer is absent and the mirror
of the FP-cavity is attached directly to the bar, i.e. to the mass M.
The corresponding equations of motion can be reduced
from (1) putting $m=0,\,H_{2}=0,\,F_{fl2}=0$.\par
The input noise spectral density will look as
$$
<|F_{n}(j\Omega)|>= 2k_B
TH_{1}\left[1+(\Omega/\Omega_{r})^{2}\frac{}{}\right], \eqno(17)
$$
where
$$
\Omega_{r}^{2}=\frac{2k_B TH_{1}}{4\pi
M^{2}\omega_{0}^{2}N_{e}}=\frac{k_B T}{M\omega_{0}^{2}}
\frac{\omega_{0}}{2\pi QN_{e}};\;
\Omega=\omega_{0}-\omega,\;|\Omega|\ll\omega_{0}.\eqno(18)
$$
(to derive this formula,  a usual approximation, typical for the
near resonance zone,  was used :
$[(\omega^{2}-\omega_{0}^{2})^{2}+4\delta^{2}\omega^{2}]
\approx 4\omega_{0}^{2} (\Omega^{2}+\delta^{2})$, $\delta\ll|\Omega|\ll\omega_{0}$).\par
Looking at the noise spectrum (17) one can conclude that the
potential sensitivity  (16) will be kept around resonance
frequency $\omega_{0}$ in the bandwidth $\pm \Omega_{r}$ defined
by the ratio of intensities of the thermal and optical (read out)
noises.

\subsection{Real sensitivity of the total scheme}

In practice, a DT is used to increase an amplitude of the detected
displacement and therefore, beside the read out noise, also the thermal
noise of DT  will limit the sensitivity. To estimate it one can use the
general formula for input noise spectrum (5). The special interest
of our calculation as in above consists in estimation of the
frequency band inside of which the sensitivity would be kept on its potential
level (16).\par Using formulas (14), (5) and (2) one can get the following
result for the input spectral noise density
$$
\begin{array}{c}
  \left<|F_{n}(j\Omega)|^{2}\right>= 2k_B
TH_{1}\Gamma(\Omega);\; \\
\\
  {\displaystyle \Gamma(\Omega)\approx
1+\frac{(\Omega^{2}-\Omega_{B}^{2})^{2}}{\omega_{r}^{4}}
+4\varepsilon^{-1}\frac{Q_{1}}{Q_{2}}\left(\frac{\Omega+\Omega_{0}}{\omega_{0}}
\right)^{2}.} \\
\end{array}  \eqno(19)
$$
This expression for the noise factor \, $\Gamma(\Omega)$ is
approximately valid in a region near the resonant frequency
$|\Omega|\ll\omega_{0}$. A specific parameter $\omega_{r}^{2}=
\omega_{0}\Omega_{r}/2 =\omega_{0}^{2}(\Omega_{r} /2\omega_{0})$
presents a normalized value of the bandwidth $\Omega_{r}$. Other
symbols in (19) are:
$\Omega_{0}=\omega_{0}-\omega_{1},\;|\Omega_{0}|\ll\omega_{0}$ is
the initial detuning and $\mu^{-1}(Q_{1}/Q_{2})\approx
H_{2}/H_{1}$ is the ratio of quality factors of the bar $Q_{1}$
and DT $Q_{2}$ associated with the corresponding losses.

In analogy with the points A, B above, one can write down the expression
for SNR using the noise spectrum (19). The potential sensitivity will be
kept only in the frequency zone where the two last terms in eq.(19) do
not exceed unity. Without loss of  generality,  it is convenient to consider
a very practical case of equal partial
frequencies $\omega_{1}=\omega_{2}$. Then the initial detuning is
absent: $\Omega_{0}=0$. Thus, requiring that the excess noise
factor $\Gamma^{+}(\Omega)=\Gamma(\Omega)-1$ be small leads to the
inequality
$$
\Gamma^{+}(\Omega)=\frac{(\Omega^{2}-\Omega_{B}^{2})^{2}}{\omega_{r}^{4}}
+4\mu^{-1}\frac{Q_{1}}{Q_{2}}\left(\frac{\Omega}
{\omega_{0}}\right)^{2}\leq 1 \eqno(20)
$$
It is interesting to note that (20) can be presented in an
oscillatory form:
$$
\Gamma^{+}(\Omega)=\omega_{r}^{-4}|K_{e}(j\Omega)|^{-2},\;K_{e}(j\Omega)=
(\Omega_{B}^{2}-\Omega^{2}
+2j\gamma_{e}\Omega)^{-1},
$$
where $K_{e}(j\Omega)$ is the transfer function of a low-frequency
equivalent oscillator having the resonance at the beating
frequency $\Omega_{B}$ with the quality factor
$$
Q_{e}=\frac{\Omega_{B}}{2\gamma_{e}}=\left(\frac{\Omega_{B}}{\omega_{r}}
\right)^{2}\sqrt{\frac{Q_{2}}{Q_{1}}}.
$$
Such new description of the excess noise factor $\Gamma^{+}(\Omega)$ allows
to perform analytically a choice of key parameters of the problem adjusting
the best "sensitivity- bandwidth" relation. \par
There are two different types of behavior of the system: the so called
"oscillation regime"\, for $Q_{e}>1/2$ and the "relaxation
regime": $Q_{e}<1/2$. The difference in these two regimes is reflected on
the character of "zones of effective sensitivity".

A solution of eq.(20) lets us define the frequency range of such
zones. It is convenient to use the following dimensionless variables:
$$
x=\left(\frac{\Omega}{\Omega_{B}}\right)^{2},\;\xi=\left(\frac{\omega_{r}}
{\Omega_{B}}\right)^{4},\; 2\nu=\left(\frac{Q_{1}}{Q_{2}}\right).
\eqno(21)
$$
then the inequality (20) is read as
$$
(x-1)^{2}+2\xi\nu x-\xi\leq 0.
\eqno(22)
$$
with roots
$$
x_{1,2}=(1-\xi\nu)\pm\sqrt{D},\,\,\,
D=(1-\xi\nu)^{2}+\xi-1,\,\quad(\xi,x_{1,2}>0), \eqno(23)
$$
and the evident constraint on values of free parameters $\xi$,\,\, $\nu$
$$
D\geq 0:\;\xi\geq(2\nu-1)/\nu^{2}\;.\eqno(24)
$$
As one can see from (23) the determinant $D$ is sensitive to
variations $\xi$ around the threshold $\xi=1$ separating two different type of solutions.

a). The "relaxation regime"\, $\xi>1$. \par
There are two real roots of the equation (22) but only the
positive one is acceptable. Then the inequality (22) is fulfilled
in the one central zone of "potential sensitivity"
$$
0\leq x \leq(1-\xi\nu)+\sqrt{D}.
$$
or coming back to frequencies (21) one estimates the width of this zone as
$$
\Delta\Omega\leq2\,\Omega_{B}\sqrt{(1-\xi\nu)+\sqrt{D}},\eqno(25)
$$

b). The "oscillation regime" \, $\xi<1$. \par
Positive roots of the equation (22) exist only if $\xi<1/\nu$ (23)
(we remark that eq. (24) is contradicted if $\nu>1$, so such regime is
possible only for $\nu < 1$). In this case a solution of the
inequality (22) results in
$$
(1-\xi\nu)-\sqrt{D}\leq x\leq(1-\xi\nu)+\sqrt{D}.
$$
It means there are two permitted frequency zones symmetrical with respect
to the central frequency $\omega_{0}$ with a total width given by
$$
\Delta\Omega\leq\Omega_{B}\frac{4\sqrt{D}}{\sqrt{(1-\xi\nu)+\sqrt{D}}
+\sqrt{(1-\xi\nu)-\sqrt{D}}}. \eqno(26)
$$
In the limiting case $\xi \rightarrow 1$ two back side zones are transformed
into the single central one (25).

\section{Numerical estimates}

Now we can return to the main target of our analysis - an estimation of
the resonant antenna bandwidth in which the sensitivity might be kept at
the potential level, limited only by the bar thermal noise.\par
Our interest to get such estimate for two variants of bar
antennas: first, the room temperature bar with DT and
FP-optical read out (Russian project OGRAN); second, a
supercryogenic bar with DT using FP-optical read out instead of
the SQUID sensor.\par
Preliminarily, we have to numerically define the level of potential
sensitivity in term of spectral noise density for both cases of
interest according to eq. (16). Typical parameters of bars are the
following: the effective length and mass --- $L\approx
2\,m,\,M\approx 10^{3}kg$, the resonance frequencies ---
$\omega_{0}=8\cdot10^{3} s^{-1}$ (OGRAN) and $\omega_{0}=5.8\cdot
10^{3}s^{-1}$ (Nautilus), the temperature $T=300$K (OGRAN) and
$T=10^{-1}$K  for Nautilus (although recent results on the
Minigrail detector \cite{grail} show that $T=10^{-2}$K is
attainable), the quality factors --- $Q=1.6\cdot10^{5}$ (OGRAN)
and $Q=6\cdot10^{6}$
(Nautilus).\par
Substitution of these values into eq. (16) leads to estimates  of the
"potential sensitivity"\\
$\alpha)$ for room temperature detector (OGRAN):
$$
|h(f)|_{n}\approx1.5\cdot10^{-20}\mathrm{Hz}^{-1/2},\;h_{\min}\approx
4.5\cdot10^{-19}. \eqno(27a)
$$

$\beta)$ for super cryogenic detector (Nautilus):
$$
|h(f)|_{n}\approx 10^{-23}\mathrm{Hz}^{-1/2},\; h_{\min}\approx
3\cdot10^{-22}. \eqno(27b)
$$
In both cases $h_{\min}$ was estimated in the the signal band:
$\Delta f = \tau^{-1}_{s}=10^{3}$ Hz, as we are dealing with "potential
sensitivity".

We then introduce the effect of readout. Assuming for the optical
readout system the following parameters: external infrared laser power
$P=1$ W ($\lambda=1 {\mu}m)$  and finesse of FP resonator
$\mathcal{F}\approx\pi/(1-R)=3000$,  one can calculate the admitted
frequency bandwidth of corresponding gravitational antenna.
\\

i) \textit{Room temperature opto-acoustical bar with DT (OGRAN)}.

At room temperature it is technically difficult to get equal
quality factors $Q_{1}\approx Q_{2}$. In practice quality factor
of the DT is usually much less the bar one $Q_{2}\ll Q_{1}$
\cite{auriga} and so $\nu=(Q_{1}/Q_{2})\gg1$. It means a such
detector can operate only in the "relaxation regime"\, $\xi>1$.
Then, taking a limit of (25) under the condition $\xi\nu \gg 1$ we
come to the estimate of effective detection bandwidth
$$
\Delta\Omega\leq2\,\Omega_{B}\sqrt{1/2\nu}=2\Omega_{B}\sqrt{Q_{2}/Q_{1}}=
\omega_{0}\sqrt{\mu Q_{2}/Q_{1}} \eqno(28)
$$
Thus the effective bandwidth $\Delta\Omega$ is defined by the
parameter of losses $\nu=(Q_{1}/2Q_{2})\gg1$ and the beating
frequency $\Omega_{B}$, i.e. the main limiting role belongs to the
thermal noise of DT. However this regime can be realized only with
a good optical sensor with sufficiently small read out noise.
Indeed, the beating frequency $\Omega_{B}$ here is limited by the
condition $\xi>1$ or in the equivalent form
 $\Omega_{B}<\omega_{r}=\sqrt{\omega_{0}\Omega_{r}/2}$, where (see
 (18), (14))
$$
\Omega_{r}^{2}=\left(\frac{\pi \mathcal{F}}{\lambda}\right)^{2}
\frac{k_B T }{M\omega_{0}^{2}}\,\frac{\omega_{0}}{2\pi
Q_{1}}\,\frac{\eta P_{0}}{\hbar\omega_{n}}\,. \eqno(29)
$$
In this formula the finesse of FP-sensor was taken as
$\mathcal{F}=2\tau^{-2}$. By substituting the values $\mu\simeq
10^{-2}$ and $Q_{1}/Q_{2}\simeq 30$ we come to the estimate:
$\Delta\Omega\simeq 0,02\omega_{0}$.
\\

 ii) \textit{Cryogenic detector with optical read out.} \par
At low temperature the DT quality factor can be large enough,
$Q_{2}\sim Q_{1}, \nu=1/2$. Therefore, the estimates of effective
bandwidth for two cases of interest take the form:
$$
\xi>1:\;\Delta\Omega\leq2\,\Omega_{B}\;;\; \eqno(30a)
$$
$$
\xi<1:\;\Delta\Omega\leq\,2\Omega_{B}\,\frac{\xi}{1+\sqrt{1-\xi}}
\eqno(30b).
$$
Analysis of the formulae (28),\,(30a),(30b) shows that: \par

i). A maximum value of the detection bandwidth $\Delta\Omega$
takes place in the "relaxation regime"\,  and it is defined by the
beating frequency $\Omega_{B}$. Optimal choice of parameters
corresponds to the condition
$$
\Omega_{B}=0.5; \quad\sqrt{\mu}\omega_{0}\approx\omega_{r};  \quad\xi\sim1,
 \quad \Omega_{r}=0.5\mu\omega_{0}. \eqno(31)
$$
These relations define the requirements for the parameters of the optical
sensor
 ($P_{0},\,\mathcal{F}$) for a given transformer factor
 $\mu=m/M$.\\

ii).In the "oscillation regime"\,  $(\Omega_{B}>\omega_{r})$ the
effective detection bandwidth is decreased in the factor
$\left(\,1+\sqrt{1-\xi}\,\right)/\xi$ but it is not too important
for the practical value of $\xi\sim 1$.

An estimate of the effective detection bandwidth for a cryogenic bar
with the optimal optical read out yields:
$\Delta\omega=\sqrt{\mu}\omega_{0}\simeq 0.1\,\omega_{0}$ i.e. the
potential sensitivity of a bar with resonance frequency close to 1
kHz can be realized in the frequency region on the order of hundred
Hz. Typical frequency bandwidth variations versus $\xi$, $\nu$
parameters are illustrated by \textit{fig.3}; the left column
corresponds to the room temperature setups, the right column to the
cryogenic ones.

\section{Discussion and conclusions}

Our analysis has shown that a room temperature bar detector
equipped with a DT with lower quality factor can actually achieve
the limit sensitivity (limited only by the thermal noise of the bar)
despite the presence of dominant DT thermal noise.
The physical reason of this is the effect of dynamical damping of DT
noises in a narrow region near the partial bar frequency
$\omega_{0}$. The depth and width of the damping is proportional to
the bar quality factor and to the coupling parameter $\mu$. However,
the width of such damping or the "width of potential sensitivity
zone"\,  is invariably smaller than the width that the same zone would
have if no  DT were used (see fig.3). A correspondent condition can
be written as
$$
\frac{Q_{2}}{Q_{1}}\leq
\left(\frac{\Omega_{r}}{\Omega_{B}}\right)^{2} \eqno(31)
$$
If the inequality (31) is fulfilled there is no advantage to use DT.

If we use the parameters of the OGRAN project: $P=100$ mW, $T=300$ K,
$\mathcal{F}= 3\cdot10^{3}$, $Q_{1}=1.64\cdot10^{5}$,
$Q_{1}/Q_{2}=30$, $\mu=10^{-2}$, $\eta=0.5$, the bandwidth defined
by the optical FP sensor is $2\Omega_{r}\sim 2.8\cdot 10^{2}$
rad/s. Application of the corresponding DT reduces the bandwidth
to $\Omega_{r}\sim 1,5\cdot10^{2}$ rad/s, i.e. results
in almost halving it. \par
As an example, consider the case of the test experiment with the optical
bar prototype of the AURIGA group \cite{auriga}:  $P=1.5 mW$,
$\mathcal{F}=2.8\cdot10^{4}$, $Q_{1}=1.8 \cdot10^{5}$, $Q_{1}/Q_{2}=27$,
 $\omega_{0}=5.45 \cdot10^{3}$ rad/s, $\mu=1.7 \cdot10^{-3}$;
 the reduction would consist of almost one order of magnitude.
The bandwidth of potential sensitivity achieved in the experiment
\cite{auriga} $\sim 43$ rad/s with DT and a short FP cavity ($\sim2$ cm)
could be expanded up to 400
rad/s without DT. But in this case the optical read out should be
realized through a long FP cavity with mirrors attached to the
bar ends. Just this technical ability,- a central tunnel along the
bar axis- is foreseen in the OGRAN project.\par
The realistic value of the bandwidth of potential sensitivity
planned for the OGRAN project \cite{cambridge} consists of a few
dozens Hertz.

For a supercryogenic bar with the physical temperature $T=10$ mK
and a high quality DT $Q\approx10^{7}$ the potential spectral
noise density $\approx 10^{-23}\mathrm{Hz}^{-1/2}$ can be achieved
in the bandwidth a hundred Hz with optical sensor of moderate
parameters $ P_{0}\simeq (0.1-1.0) W $, $\mathcal{F}\sim
(10^{4}-3\cdot 10^{3})$.

A more conservative estimate of sensitivity $\sim 3\,10^{-21}\mathrm{Hz}^{-1/2}$
and bandwidth $50\,\mathrm{Hz}$ for cryogenic bar with a small gap optical
 sensor operated as an accelerometer can be found in ref.\cite{auriga2}.
However it was limited by technical noises described in that paper.
In particular the additive technical laser power noise was included and
it was taken a factor 200 larger than DT thermal noise. The temperature
$T=0.1$ $K$ was 10 times larger than  our estimation above. Finally,
the product of important parameters, crucial for increasing sensitivity
$PL_{t}^{2} \mathcal{F}^{2}$ (laser power, DT length, DT cavity finesse)
was a factor 40 less. So the total noise power gain turns out to be of
the order of $10^{4}$ in favor of our scheme. It results in a factor ~300
for the sensitivity estimate in term of metric perturbation. The frequency
spectral density noise estimation has the same order of value in both cases.
Other technical noises of a PZ-actuator and its driver used in the paper
\cite{auriga2} did not exceed the frequency noise (with the parameters of
the AURIGA setup). \par
It worth to remark here the advantage of the FP sensor with a small cavity
gap is known as a method of suppressing  laser technical frequency fluctuation.
Regarding  laser power fluctuation, the situation is opposite: a corresponding
frequency noise (induced by power variations) is inversely proportional to
the cavity length. In that respect, an OGRAN cavity as long as the gravitational
bar scale looks preferable.\par
In this paper we only focused on fundamental natural fluctuation: thermal
mechanical noises and optical short noise, neglecting technical noises. However,
the Virgo and Ligo experiments have proven that technical noises
of the laser pump, potodetectors, mirror drivers etc. can be sufficiently reduced,
at least in the kilohertz frequency region. so that the light power fluctuation
foreseen in the Virgo interferometer does not exceed the corresponding shot noise
level $|<\Delta P_{f}^{2}>|^{1/2} \sim 10^{-10}\,W\mathrm{Hz}^{-1/2}$ \\
We believe that further increases of laser power and mirror finesse will be
available in the near future. This would allow to further expand the detection
bandwidth up to $10^{3}$ Hz (27b)(in our analysis we could not calculate it due
to our restriction of "close to resonance approximation"\, (17),(19)).

One serious problem we encounter on this way is matching the optical cavity with
power pump when dealing with mirrors cooled to very low temperatures. In fact,
at present this problem has not yet a definite experimental solution. Investigation
of cooled mirrors with incident and reflected optical power on the order of 1 W
(and more) is carried out now by interferometer groups as advanced setups working
at cryogenic temperature are planned. As an example, the well known Japanese LCGT-project
 \cite{LCGT} is designed for operation with mirrors cooled at  $20 K$
and 100 W light power. Preliminary experiments have shown that mirrors with absorption
1 ppm and power 0.1W cooled to 4 K do not increase a mechanical and optical noise
level typical for present generation of interferometers. There are indeed no experiments
with mirrors at supercryogenic temperature. However we believe that the thermal
contact of the mirrors with cryostat in the case of the "bar-interferometer" will be
much better than for suspended mirrors (free mass interferometer).
Recent advances in dilution refrigeration technology have made  commercially available
\cite{LC} devices with large cooling power at very low temperatures (up to $50 \mu W$ at 30 mK).
So, the problem of mantaining the low temperature of mirrors illuminated by the light
beam will probably reduce to  proper design of the cryostat.
\par
Our calculation above was performed also under the simplification
of "zero detuning"\, $\omega_{0}=\omega_{1}$. Non perfect tuning
produces a mismatch and in principle destroys the balance of
noises. But if such detuning does not exceed the beating frequency
all results formulated above remain valid. \par
It worth to remark that the reception band of GW bars might be expanded
through a parametrical regeneration of the DT oscillator. Theoretically
this possibility was considered in the paper \cite{guru}. The electrical rf-resonance circuit
$(\omega_{0})$ coupled with the DT having a DC bias can be regenerated by applying also
AC pump at the double frequency $(2\omega_{0})$. The regeneration energy is regulated by
alternative amplitude of the pump at some so called "frequency of superization"
$\Omega_{sp}\ll\omega_{0}$ (see details in \cite{guru}). Finally,  it leads to an
equivalent increase the reception bandwidth and transfer function. However the application
of this method to the DT with optical FP cavity has to be specifically addressed.

As a further development, we are considering with interest a
set-up with each mirror of the FP mounted on a separate DT, thus
fully exploiting  the long baseline feature of this readout. In
this case, the so called {\it Janus} scheme \cite{giano}, we deal
with a 3 mode mechanical system (bar +two DTs on opposite end
faces) that has also very interesting symmetry  properties with
respect to a real g.w. signal : indeed, of the three resulting
normal modes, the central one has no quadruple moment, and can
then be used as an effective internal veto.   Further analysis,
needed to characterize this setup, is in progress.

\section*{Acknowlegments}

The authors would like gratitude their colleagues professors
G.Pizzella , G.Pallottino, F.Ricci (Universities of Roma), E.Majorana (INFN)
and also S.Vyatchanin, V.Mitrofanov (MSU) for useful discussions while
preparing this paper.

\section*{Appendix}

A critical magnitude of laser power is defined by an equality of
the thermal noise of mirrors and its fluctuation under the light
photon pressure. The pressure force is described as stochastic
series of photon shots
$$
f(t)=(1+R)\frac{\hbar\omega_{n}}{c}\,\sum\limits_{\nu}\delta(t-t_{\nu})
$$
which is considered as the Poissonian process with moments
$$
\left<f(t)\right>=\frac{\hbar\omega_{n}}{c}N_{1},\;
\left<\widetilde{f}(t)\widetilde{f}(t+\tau)\right>=
\left(\frac{\hbar\omega_{n}}{c}\right)^{2}N_{1}\delta(\tau).
\eqno(A1)
$$
Here $N_{1}=P/\hbar\omega_{n}$ is the average rate of photons
inside the FP cavity; $P=P_{0}/(1-R)$, $P_{0}$ is the external
laser power.

Thus the correlation function of the light pressure force read as
$$
\left<\widetilde{f}(t)\widetilde{f}(t+\tau)\right>=(1+R)^{2}\left
(\frac{\hbar\omega_{n}}{c}\right)^{2}N_{1}\delta(\tau)
\approx4\frac{\hbar\omega_{n}P_{0}}{(1-R)c^{2}}\delta(\tau).
$$
It has to be compared with the Nyquist force correlation function
$\left<f_{T}(t)f_{T}(t+\tau)\right>=2k_B TH\delta(\tau)$.
Finally the critical laser power is given as
$$
[P_{0}]_{\mathrm{\textrm{cr}}}=4\left(\frac{k_B
T}{\hbar\omega_{n}}\right)\,\frac{m\omega_{0}}{Q\mathcal{F}}\,c^{2}.
$$
Substitution typical parameters: $T=10$ mK, $\lambda=1 \mu$m, 
$m=10^{4}$g, $\omega_{0}\approx6\cdot10^{3}$ rad/s,
$Q\approx10^{7}$, $\mathcal{F}\approx10^{4}$ results in the
estimation $[P_{0}]_{\mathrm{cr}}\approx160\cdot10^{3}$ kW i. e.
too large a value for a practical laser pump.

\begin{figure}[p]
\makebox[\textwidth][c]{
\includegraphics[width=15cm]{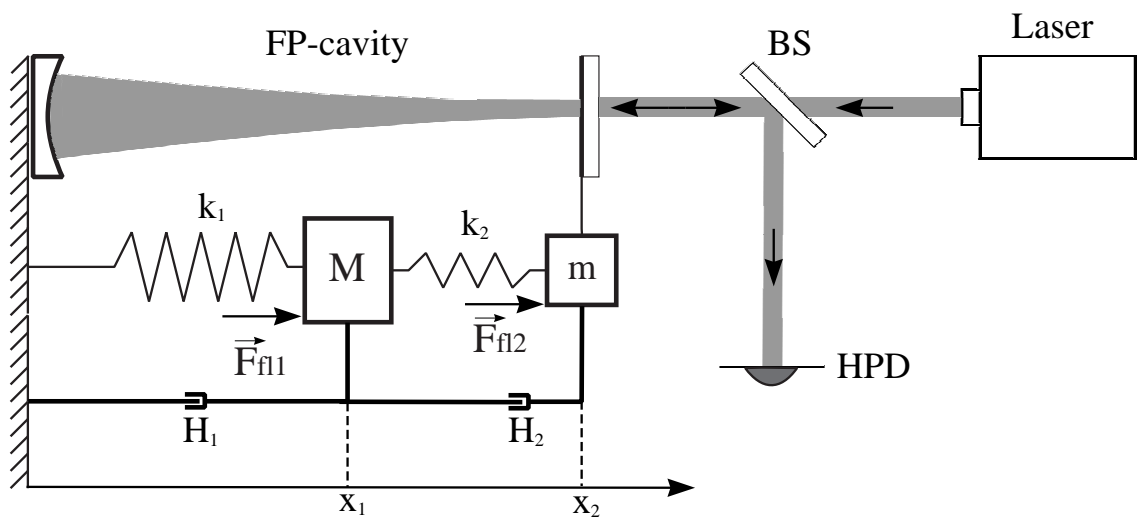}}
\caption{Equivalent scheme of the opto-acoustical detector. $M,
k_{1}, H_{1}$ -- mass, rigidity and friction coefficient of the bar.
$m, k_{2}, H_{2}$ -- correspondent transducer parameters.
}
 \label{f:1-1}
\end{figure}

\begin{figure}
\makebox[\textwidth][c]{
\includegraphics[width=15cm]{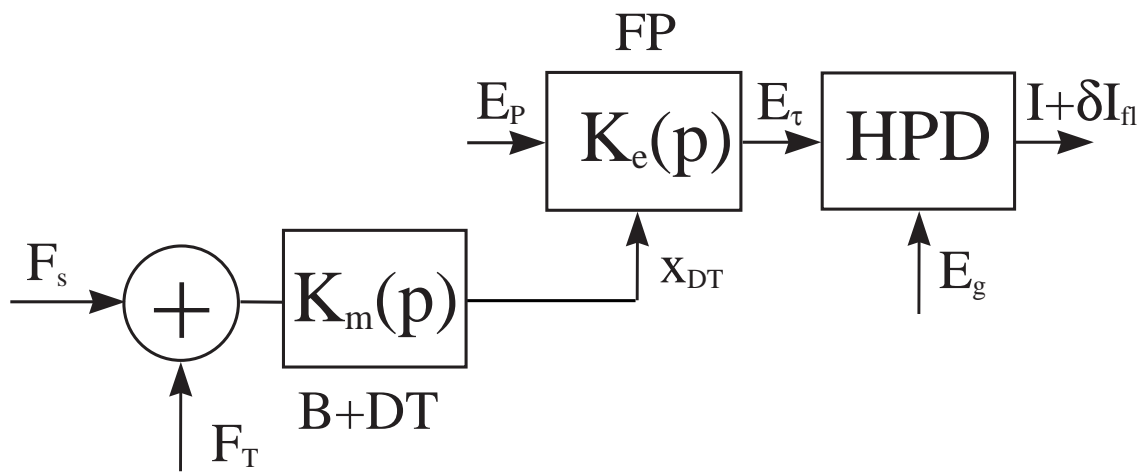}}
\caption{Block diagram of the signal/noise track. $K_{m}(p),
K_{e}(p)$ -- mechanical and optical transfer functions of the setup.
HPD = a heterodyne photo detector}
 \label{f:1-2}
\end{figure}

\begin{figure}
\makebox[\textwidth][c]{
\includegraphics[width=\textwidth]{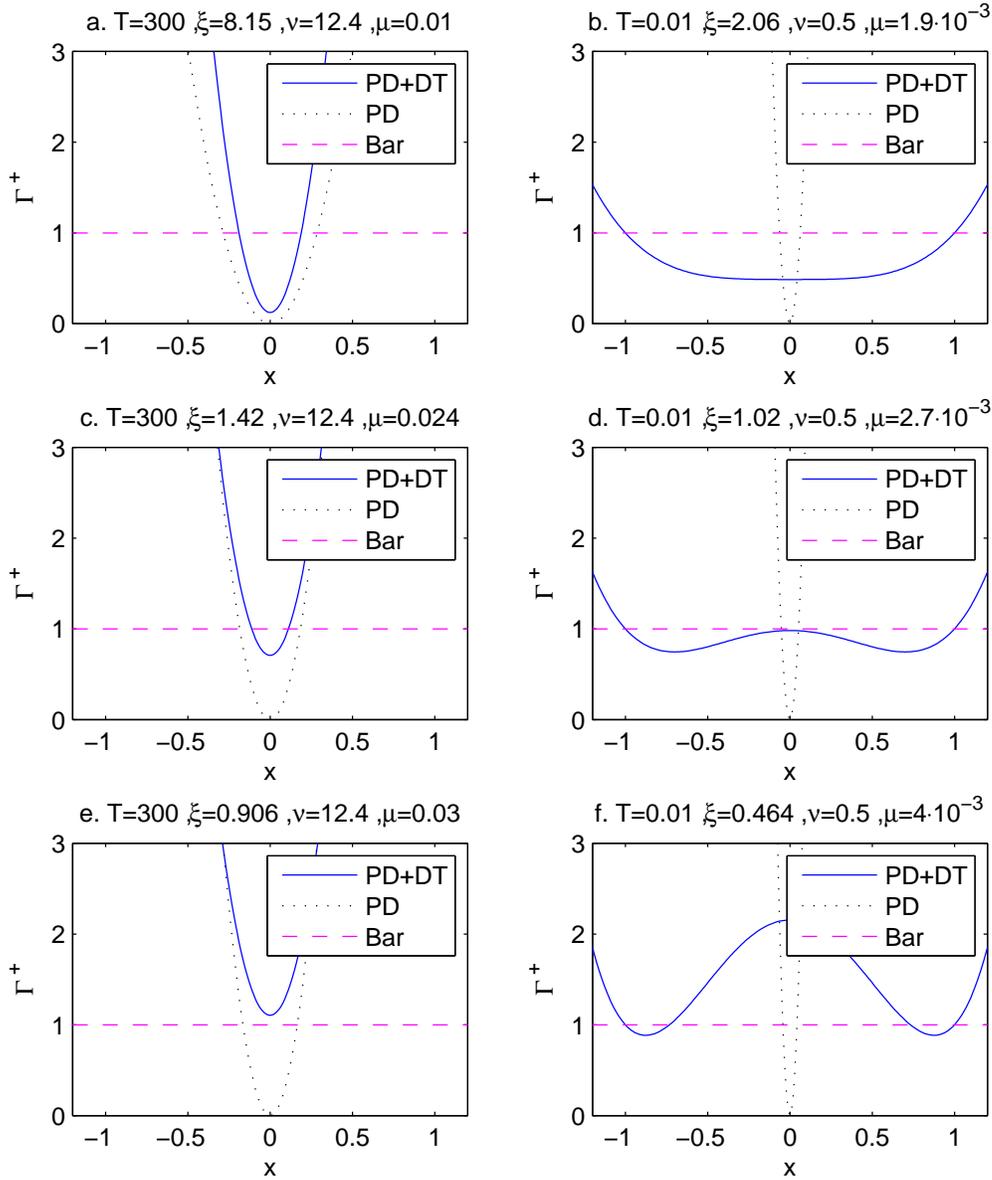}}
\caption{Budget of noises $\Gamma^{+}=<|F_{n}(j\omega)|> 2k_B
TH_{1}$ versus parameters $ \nu, \xi $($\nu=Q_{1}/2Q_{2}$;
$\xi=\omega_{r}/\Omega_{B}$), $x=\Omega/\Omega_{B}$ PD-optical readout only, PD+DT -
optical readout with transducer. Dashed line is bar noise}
 \label{f:1-3}
\end{figure}
\end{document}